\documentstyle[pra,aps,amsfonts,]{revtex}

\begin{document}

\draft
\twocolumn[\hsize\textwidth\columnwidth\hsize\csname 
@twocolumnfalse\endcsname

\title{Dimensional crossover in the fluctuation magnetization in $YBa_2Cu_3O_{7-x}$ and its evolution with the content of oxygen}
\author{S. Salem-Sugui Jr.$^{1\ast}$, A. D. Alvarenga$^2$, B. Veal$^3$\\ and A. P. Paulikas$^3$}
\address{$^1$ Instituto de Fisica, Universidade Federal do Rio de Janeiro, 21945-970 Rio de Janeiro, Brasil\\
$^2$ Centro Brasileiro de Pesquisas Fisicas, Rua Dr. Xavier Sigaud 150, Rio de Janeiro, Brazil.\\
$^3$ Materials Science Division, Argonne National Laboratory, Argonne, IL 60439, USA}

\date{\today}

\maketitle
\begin{abstract}
We have studied the field-induced dimensional crossover in the fluctuation magnetization of three single crystals of $YBa_2Cu_3O_{7-x}$, with superconducting transition temperatures, $T_c$= 62.5, 52, and 41 K. The dimensional crossover is observed by studying the diamagnetic vortex fluctuations of the lowest-Landau-level type which occur in isochamps magnetization curves, $MvsT$, for temperatures close to the transition temperature $T_{c}(H)$. The study was accomplished by obtaining isochamps magnetization curves as a function of temperature for fields in the range of 0.1 T to 5 T. Magnetization curves for each sample when plotted together show two distinct well resolved crossing points, one formed by low field curves and located at a higher temperature than the other formed by high field curves. A lowest-Landau-level scaling analysis is applied to the curves forming the crossing points  and it is verified that lower field curves obey the three-dimensional form of this scaling while the higher field curves obey the two-dimensional form. The results allow to observe the evolution of the dimensional crossover field $H_{cross}$, 3D-2D, with the content of oxygen in $YBa_2Cu_3O_{7-x}$. It is observed that the evolution of the field $H_{cross}$ with the content of oxygen in each sample qualitatively agrees with theoretical predictions and allow us to estimate the ratio of the anisotropy among the studied samples. \\pacs{74.40.+k}
\end{abstract}
]

Large thermal fluctuations have been observed in high-$T_c$ superconductors (HTS) near the upper critical field $H_{c2}$ \cite{Lee,ullah,klemm,welp,tesa1,bula,tesa2,said1,said2,moloni,rosenstein}. These fluctuation effects are larger than observed in conventional low-$T_c$ superconductors \cite{said3}, due to the HTS's short coherence lengths $\xi $ and high degree of anisotropy, $\gamma$.\\
Theoretical works have shown \cite {ullah,klemm} that in the
presence of strong magnetic fields, for which the lowest-Landau-level, LLL,
approximation can be used, the $\Delta T$ fluctuation region around $T_{c}$, for
a system with dimensionality $D$ is proportional to $(TH)^{(D-1)/D}$, and
that within this
region physical quantities such as magnetization and direct-current conductivity should be scaled with $(TH)^{(D-1)/D}$. For magnetization in
particular, scaling predicts that $MvsT$ data obtained at
different fields, $H$, should collapse onto a single curve when the variable 
$M/(TH)^{(D-1)/D}$ is plotted against $(T-T_c(H))/(TH)^{(D-1)/D}$. Here, $T_{c}(H)$ becomes a fitting parameter. The LLL theory also predicts the existence of a temperature $T^*$ for which the magnetization $M(T^*)$ is independent of the applied magnetic field \cite{tesa1,bula,said2,rosenstein}. The later reflects the existence of crossing points when magnetization curves are plotted together. \\
The above scaling law has been used to
identify LLL fluctuations in a given material, and to determine its
dimensionality \cite{welp,tesa1,bula,tesa2,said1,said2,moloni,rosenstein,said3}.
It has been well established \cite{klemm} that fluctuations of a vortex lattice for layered superconductors maintain a three-dimensional, 3D, character at low fields, until a crossover field $H_{cross}$ is reached. Above $H_{cross}\sim1/\gamma^2$ \cite{glazman}, fluctuations of a vortex lattice are of quasi-2D nature. This field induced reduction of dimensionality in the fluctuation-magnetization, or dimensional crossover, has been recently observed experimentally in a sample of deoxygenated Y123 by Rosenstein et al. \cite{rosenstein}. These Authors \cite{rosenstein} observed that isochamps magnetization curves plotted together exhibit two distinct field independent crossing points. One due to the crossing of low field curves occurring at a higher temperature than the other due to the crossing of higher field curves. The Authors \cite{rosenstein} shown theoretically and experimentally that the low-field curves producing the higher temperature crossing point obey the three-dimensional-LLL scaling form while the high-field curves producing the lower temperature crossing point obey the two-dimensional-LLL scaling form. Previously, a crossing point of $MvsT$ curves due to 3D vortices fluctuations was observed in Y123 with $T_c$=92 K \cite{said2} while a crossing point due to 2D vortices fluctuations was observed in deoxygenated Y123 \cite{gomis}. \\
$YBa_2Cu_3O_{7-x}$ is the most studied high-$T_c$ material, and  deoxygenated Y123 is a system where reduction of oxygen reduces the superconducting temperature $T_c$ and increases the anisotropy \cite{veal,lifang}. Deoxygenated Y123 is then a perfect system to study the evolution of the dimensional crossover with varying anisotropy. Also, since the field $H_{cross}$ is theoretically predicted to be a function of $\sim\gamma^2$ \cite{glazman}, the study of the evolution of the field induced dimensional crossover in deoxygenated Y123 may allow a direct comparison with the theoretical predictions. The motivation of this work was to address the above issue, by extending to low fields a previous study of two-dimensional high-field diamagnetic fluctuations which was performed on deoxygenated Y123 single crystals \cite{JLTP}. The result of this study shows that the evolution of the 3D-2D crossover field observed for each sample qualitatively agrees with the theoretical predictions for a system with varying anisotropy.\\
In the present work we have obtained isochamps magnetization curves as a function of temperature in three single crystals of $YBa_2Cu_3O_{7-x}$ with critical temperatures $T_c$=62.5 ($x$=0.35), 52 ($x$=0.5) and 41 K ($x$=0.6). With the exception of the single crystal with $T_c$=62.5 K which was fully studied here, the other two single crystals were already studied for high fields (1-5T) in Ref.17, and here we only extend measurements to low fields in these two single crystals. The single crystals of Y123 were grow at Argonne National Laboratory \cite{veal}, and have approximate dimensions of 1x1x0.2mm and $mass\sim1 mg$ with the $c$ axis along the shorter direction, and each exhibited sharp, fully developed transitions ($\Delta T_{c}\simeq1K$).\\
A commercial magnetometer based on a superconducting quantum interference device (SQUID) (MPMS Quantum Design) was utilized. The scan length was 3 cm, which minimized field inhomogeneities. Experiments were conducted by obtaining isochamps magnetization, $M$, data as a function of temperature, producing $MvsT$ curves, with values of field running from 0.1 T to 5 T. Magnetization data were always taken after cooling the sample below $T_{c}$ in zero applied magnetic field, zfc (but in the presence of the Earth magnetic field). After cooling to the desired temperature, a magnetic field was carefully applied (without overshooting), always along the $c$ axis direction of the samples, and $MvsT$ curves were obtained by heating the sample up to temperatures well above $T_{c}$, for fixed $\Delta T$ increments. We also obtained several field-cooled curves, corresponding to cooling the sample from above $T_c$ to below $T_c$, in the applied magnetic
field. This procedure allowed determination of the reversible (equilibrium)
magnetization.\\
The resulting  magnetization curves $MvsT$ for each sample were plotted together, after a proper background correction. Background corrections for all samples were of the type $M(T)$ = $A$ + $B/T$, where $A$ and $B$ are constants determined by fitting a selected region of a given curve well above $T_{c}(H)$. An eye inspection of the plot of the $MvsT$ curves for each sample show the existence of two well-defined crossing points of these curves. One crossing point is formed by low field curves and occurs at a higher temperature than the other formed by high field curves. For better visualization, these crossing points are shown in the inset of the main figures where the correspondent scaling analysis is performed on the data. We proceed an analysis of the reversible magnetization curves based on the lowest-Landau-level scaling law discussed above. The scaling analysis is performed by plotting the quantity $M/(TH)^{(D-1)/D}$ against $(T-T_c(H))/(TH)^{(D-1)/D}$ where D is the considered dimension and $T_{c}(H)$ is a fitting parameter choose to produce a collapsing of the analyzed curves. The interesting result of the scaling analysis, which is presented below, is that for all samples, the crossing point formed by $MvsT$ curves obtained with low fields obey the 3D-LLL scaling law, while the crossing point formed by $MvsT$ curves obtained with high fields obey the 2D-LLL scaling law. Consistently, the crossing point produced by the curves obeying 3D vortex fluctuations occurs at a higher temperature than the crossing point produced by the curves obeying 2D vortex fluctuations. We also verify that data obeying 2D-LLL scaling does not obey 3D-LLL scaling and vice-versa. \\

\begin{figure}[htb]
\vspace*{14cm}
\special{eps: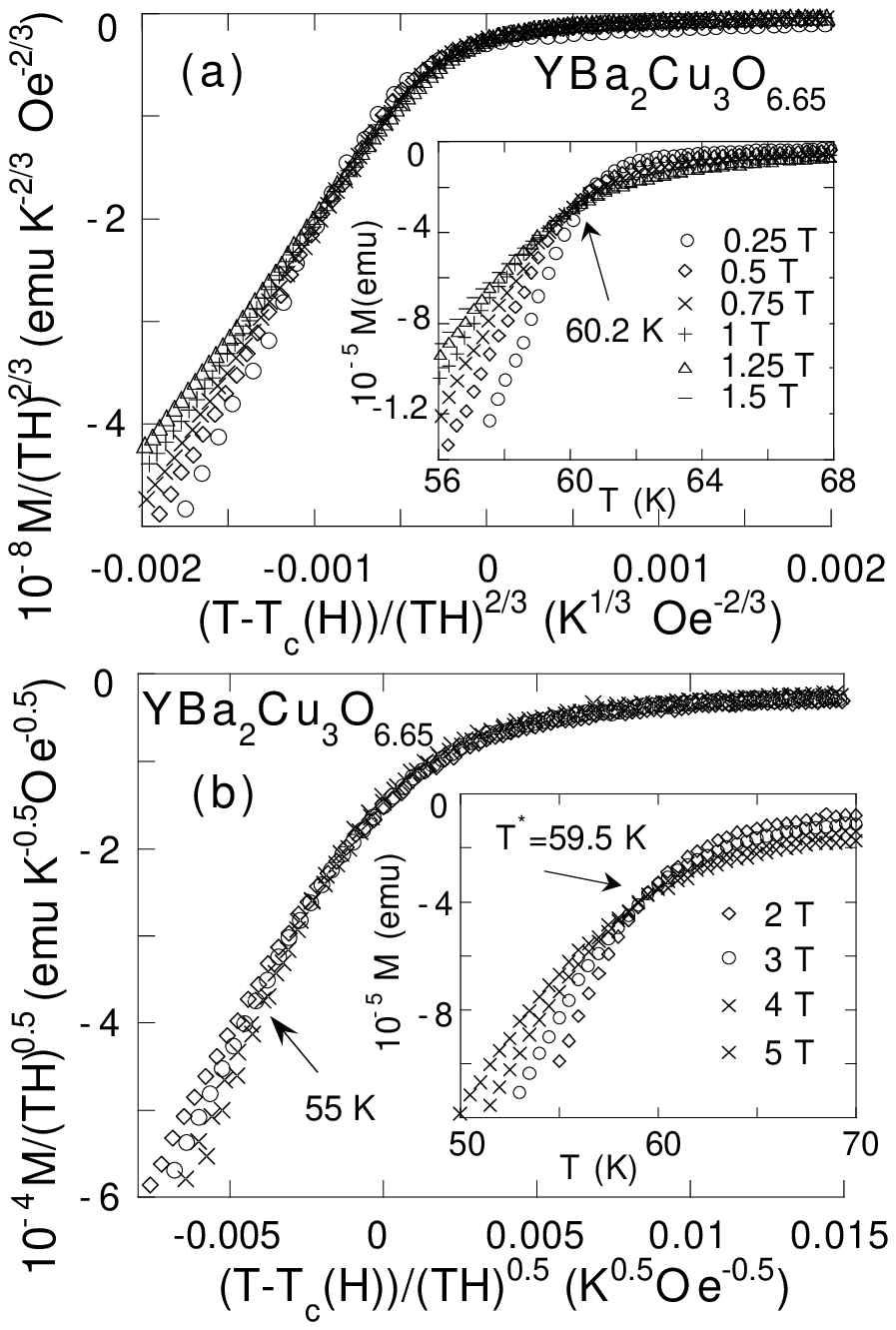 x=8.5cm y=14cm}
\caption{Fluctuation magnetization and lowest-Landau-level (LLL) for Y123 with $T_c$=62.5 K : a) Three-dimensional (3D) LLL scaling analysis. b) Two-dimensional (2D) LLL scaling analysis. The insets show $MvsT$ curves used in the main figures.} 
\end{figure}

Figures 1a and 1b show respectively the 3D-LLL and 2D-LLL scaling analysis performed on the reversible data shown in each respective inset for sample with $T_c$=62.5 K. The crossing point for data obeying 3D-LLL, corresponding to low-field curves, occurs at $T^*$=60.2 K (inset of Fig. 1a) while data obeying 2D-LLL corresponding to high-field curves, show a crossing point at $T^*$=59.5 K (inset of Fig. 1b). A comparative inspection of Fig.1a and 1b allow one to observed that the 3D-2D dimensional crossover for this sample occur for a field $1.5T \prec H_{cross} \leq 2T$. Figures 2a and 2b show the results of the analysis for the sample with $T_c$=52K, where data obeying 3D-LLL, corresponding to low-field curves, show a crossing point at $T^*$=51.2 K (inset of Fig. 2a) and data obeying 2D-LLL, corresponding to high-field curves, show a crossing point at $T^*$=50.2 K (inset of Fig. 2b). For this sample the 3D-2D dimensional crossover occur for a field $0.7T \prec H_{cross} \leq 1T$. It should be mentioned that the single crystal data corresponding to Fig. 2 has the same content of oxygen of the single crystal studied in Ref. 11 (x=0.5) and consistently we have obtained the same value of $H_{cross}\sim1 T$.

\begin{figure}[htb]
\vspace*{14cm}
\special{eps: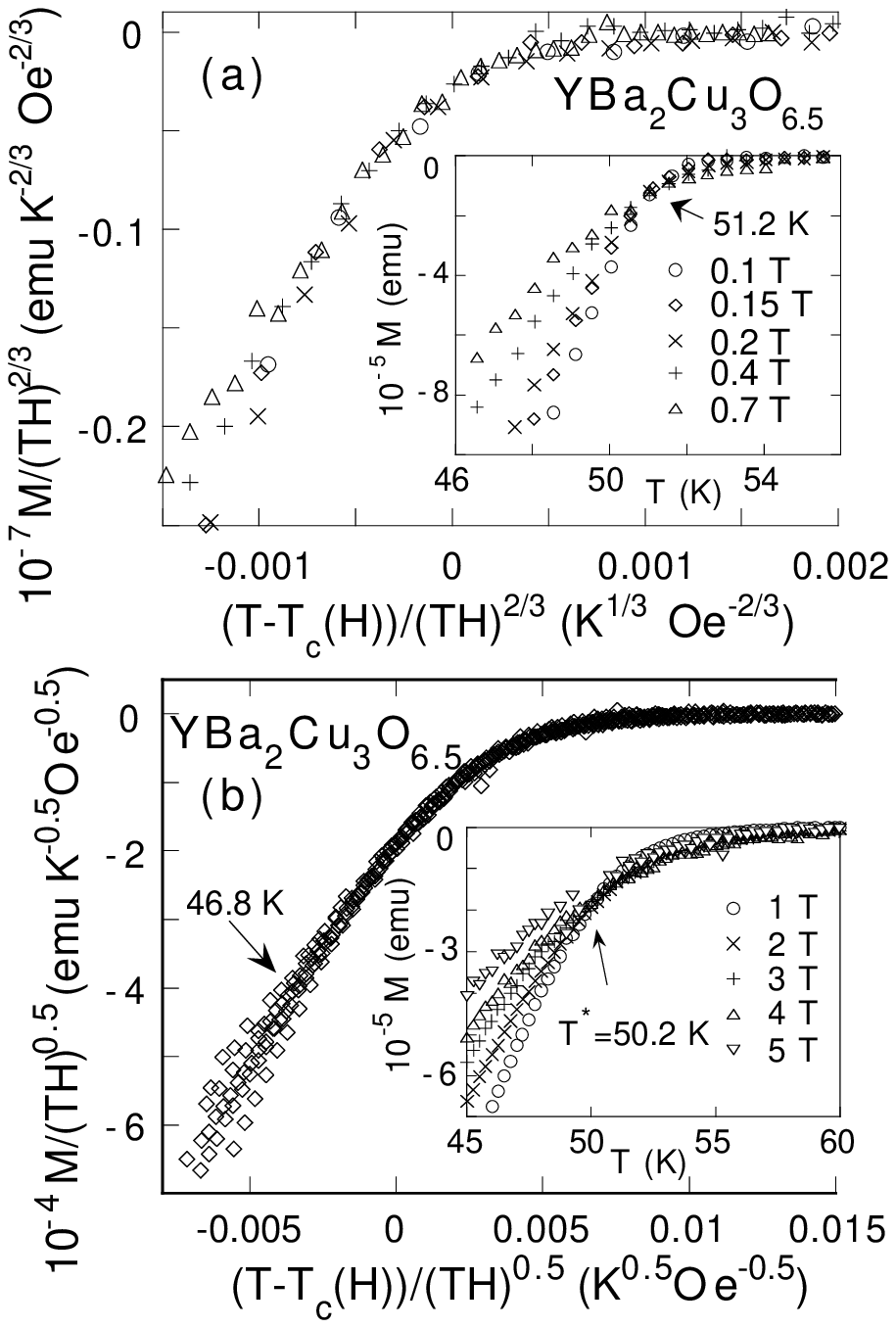 x=8.5cm y=14cm}
\caption{Fluctuation magnetization and lowest-Landau-level (LLL) for Y123 with $T_c$=52 K : a) Three-dimensional (3D) LLL scaling analysis. b) Two-dimensional (2D) LLL scaling analysis. The insets show $MvsT$ curves used in the main figures.} 
\end{figure}

Finally, figures 3a and 3b show the results for the sample with $T_c$=41 K, where $T^*$=40.4 K for data obeying 3D-LLL (inset of Fig. 3a, low-field curves) and $T^*$=39.8 K for data obeying 2D-LLL (inset of Fig. 3b, high-field curves). For this sample the 3D-2D dimensional crossover occur for $0.4T \prec H_{cross} \leq 0.5T$. 

\begin{figure}[htb]
\vspace*{14cm}
\special{eps: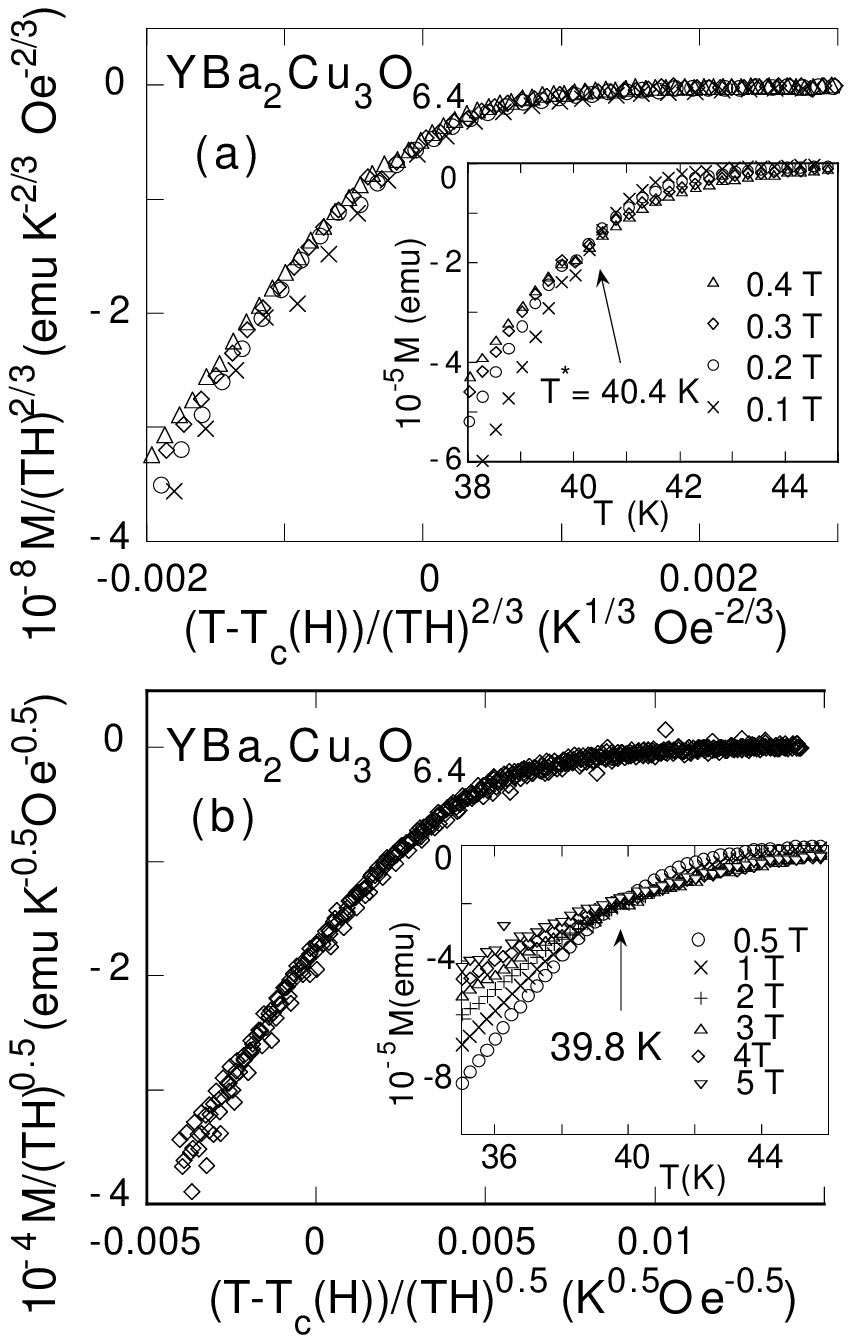 x=8.5cm y=14cm}
\caption{Fluctuation magnetization and lowest-Landau-level (LLL) for Y123 with $T_c$=41 K : a) Three-dimensional (3D) LLL scaling analysis. b) Two-dimensional (2D) LLL scaling analysis. The insets show $MvsT$ curves used in the main figures.} 
\end{figure}

To exemplify the fact that high-field data shown in Figs. 1-3 obey 2D-LLL in a wider temperature range than 3D-LLL, we show in Figure 4 the same high field data presented in Figs. 1b, 2b and 3b after scaled following the 3D-LLL scaling law. The arrow in figure Fig. 1b marks the data for the lowest field (2T) of the plot (at 55 K) above which all $MvsT$ curves collapse following 2D-LLL scaling. The same data is marked with an arrow in Fig. 4a (3D-LLL scaling analysis) and occurs in a region where $MvsT$ curves are spread apart. This fact evidences that the 2D-LLL scaling law is obeyed in a much larger temperature region for sample with $T_c$= 62.5 K. A direct comparison between Figs. 2b and 4b also shows that high-field data for the sample with $T_c$= 52 K obey 2D-LLL instead 3D-LLL. For sample with $T_c$= 41 K the high field data, show in the inset of Fig. 4b,  fail to follow the 3D-LLL scaling.

\begin{figure}[htb]
\vspace*{4cm}
\special{eps: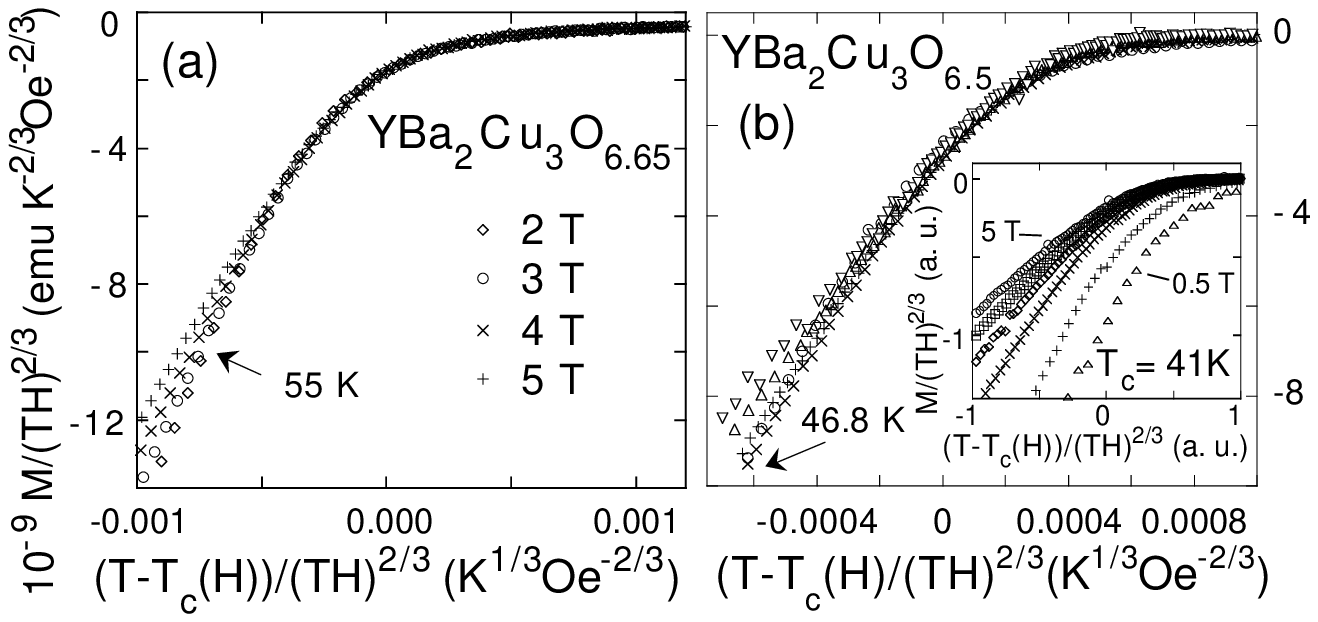 x=8.5cm y=4cm}
\caption{Three-dimensional (3D) LLL scaling analysis performed on high field data: 4a) 3D-LLL scaling performed on data of Fig. 1b. 4b) 3D-LLL scaling performed on data of Fig. 2b. The inset of Fig. 4b shows the scaling analysis after performed on data of Fig. 3b.} 
\end{figure}

It is interesting to observe that the value of the field $H_{cross}$, as obtained above, consistently decreases as the content of oxygen decreases. Or in other words, $H_{cross}$ decreases as the anisotropy increases. This effect is expected by theory \cite{klemm} and from the relation $H_{cross}\sim 1/\gamma^2$ \cite{glazman} we can estimate the ratio of anisotropy between the studied samples: $\gamma_{(T_{c}=41K)}/\gamma_{(T_{c}=62K)}\approx 2$ and $\gamma_{(T_{c}=41K)}/\gamma_{(T_{c}=52K)}\approx \sqrt{2}$. It is worth mentioning that the values of $T_c(H)$ obtained from the LLL scaling analysis are in reasonable agreement with values of $T_c(H)$ estimated for each $MvsT$ curve from the standard linear extrapolation of the reversible magnetization down to M = 0 \cite{abrikosov}. These values of $T_{c}(H)$ are plotted in Fig. 5. As it is shown in Fig. 5 all samples exhibit a linear behavior of the upper critical field in the studied range, and the arrows indicate the values of $dH_{c2}/dT$ of each sample.

\begin{figure}[htb]
\vspace*{7cm}
\special{eps: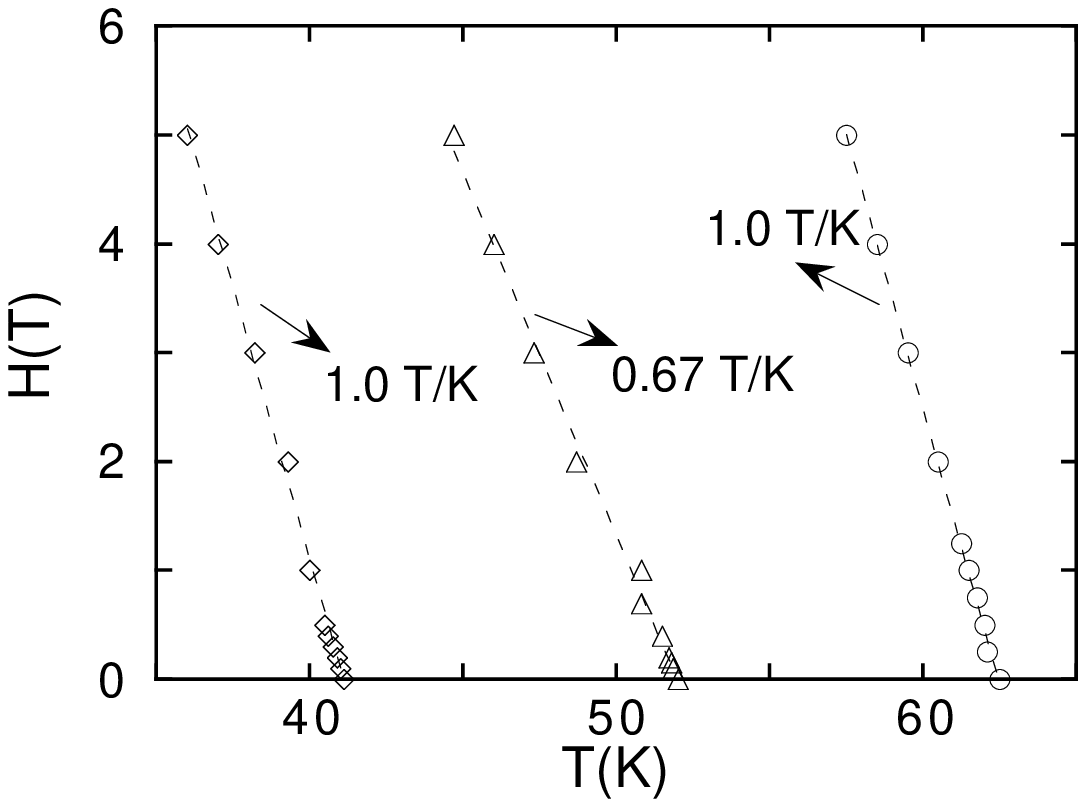 x=8.5cm y=7cm}
\caption{Magnetic phase diagram, $H_{c2}(T)$ vs $T$ for the studied samples} 
\end{figure}

In conclusion, our study of vortex fluctuations in three single crystals of deoxygenated YBaCuO show the existence of a dimensional crossover from three-dimensional to two-dimensional (reduction of dimensionality) which occurs above a certain field $H_{cross}$. It is found that the value of the field $H_{cross}$ decreases as the content of oxygen in the sample decreases (or, as the anisotropy increases), and varies from $\sim2T$ for YBaCuO with $T_c$=62.5 K to $\sim0.5 T$ for YBaCuO with $T_c$=41 K. We also observe that the evolution of the field $H_{cross}$ for the studied samples qualitatively agrees with theoretical predictions and allow to estimate the ratio of the anisotropy among the studied samples.\\
This work was partially supported by CNPq, Brazilian Agency.\\
$\ast$ Corresponding author. E-mail: said@if.ufrj.br

\end{document}